\documentclass[sn-basic]{sn-jnl}

\usepackage{graphicx}%
\usepackage{multirow}%
\usepackage{amsmath,amssymb,amsfonts}%
\usepackage{amsthm}%
\usepackage{mathrsfs}%
\usepackage[title]{appendix}%
\usepackage{xcolor}%
\usepackage{textcomp}%
\usepackage{manyfoot}%
\usepackage{booktabs}%
\usepackage{algorithm}%
\usepackage{algorithmicx}%
\usepackage{algpseudocode}%
\usepackage{listings}%

\theoremstyle{thmstyleone}%
\theoremstyle{thmstyletwo}%

\theoremstyle{thmstylethree}%

\begin{document}

\title[SPACE WEATHER PROTECTION]{Terrestrial space weather protection through human-produced mass-loading}


\author*[1]{\fnm{B.~M.} \sur{Walsh}}\email{bwalsh@bu.edu}
\author[2]{\fnm{D.~T.} \sur{Welling}}
\author[2]{\fnm{Z.} \sur{Huang}}

\affil[1]{\orgdiv{Center for Space Physics}, \orgname{Boston University}, \orgaddress{\city{Boston}, \state{MA}, \country{USA}}}

\affil[2]{ \orgname{University of Michigan}, \orgaddress{\city{Ann Arbor}, \state{MI}, \country{USA}}}


\abstract{
While humans become more reliant on Earth's space environment, the potential for significant harm from severe space weather continues to grow. As structures from the sun reach Earth's magnetosphere and space environment, they deposit energy that fuels geomagnetic storms. Currently, space weather researchers work to predict the timing and intensity of space weather events, often providing warnings of several days prior to the initiation of a strong geomagnetic storm. Here a new paradigm is presented where, rather than prediction, active steps are taken to mitigate the impact of solar wind structures through temporarily modifying Earth's magnetosphere. Global magnetohydrodynamic simulations are used to demonstrate that artificial mass-loading Earth's dayside magnetosphere can fortify Earth's space environment against strong space weather events. The simulations and supporting analysis use realistic mass-loading from model spacecraft at geosynchronous orbit to show the validity of the enabling physics as well as technical feasibility with current technology. The results demonstrate that with modern technology, the intensity of a major geomagnetic storm could be actively reduced by 50\% or more, protecting technology and human life.}




\maketitle

\section{Introduction} \label{sec:intro}
Earth resides within a dynamic region of space where conditions can impact, sometimes violently, human activity and technology in space and on the surface of Earth. This interaction is known as space weather. Often times, the most extreme space weather impacts occur during geomagnetic storms, periods of enhanced energy transport from the sun's solar wind into Earth's magnetosphere. Space weather impacts stem from a number of physical processes including increased or variable satellite drag, spacecraft charging, radiation damage and effects on electronics and astronauts, as well as geomagnetically induced currents (GICs) driven through Earth and power grids \citep{pulkkinen2017geomagnetically}.


The impact and risk of space weather or similar stochastic systems are often described as a function of occurrence frequency. A 1 in 10 year geomagnetic storm can have implications for spacecraft communication to the ground, power grids, and spacecraft operations. An example of this is the ``Halloween storm'' which occurred in 2003 and caused power outages in high-latitude locations as well as the loss of numerous spacecraft \citep{weaver2004halloween}. A 1 in a 100 year geomagnetic storm is modeled to have a number of more severe and wide-ranging damages in space and on the ground including wide-spread damage of power grids and electrical systems. The cost estimates for damages to the power grid alone are predicted to be USD 2.4 to 3.4 trillion \citep{schulte2014severe}.

On much longer scales, the impact of 1 in a million or 1 in 10 million-year event would far more devastating. Studies have linked these types of space weather events to mass extinctions on Earth \citep{lingam2017risks}. In these types of rare, but extreme geomagnetic storms, properties such as the atmosphere of a planet, are impacted to the extent that it reduces habitability. \citet{loper2019carrington} proposed that a technological species would have a narrow window of time between major space weather events in which they must either develop a defense or hardened infrastructure to withstand space weather or expand beyond the home planet. The mass-loading proposal presented here is the first human-developed defense mechanism shown to reduce the impact of a solar wind structure - an initial step in fortifying Earth against space weather risks.

\subsection{Solar wind-magnetosphere coupling}
The majority of the energy that fuels Earth's space environment to generate geomagnetic storms is extracted from the flowing solar wind. As the solar wind reaches the obstacle of Earth's magnetosphere, energy is transferred through the magnetopause boundary via magnetic reconnection. If reconnection occurs efficiently, shocked solar wind deposits energy rapidly into the magnetosphere. By contrast, if reconnection occurs inefficiently, flow patterns in the magnetosheath will be modified to divert flow around the boundary, sweeping plasma downtail and past Earth where it is lost from the system instead of reconnecting to transfer energy.

The efficiency of magnetic reconnection is governed by local plasma conditions at the reconnecting site. The efficiency depends on the plasma density and magnetic field magnitude and orientation on both sides of a current sheet \citep{cassak2007scaling}. If the mass density increases, the efficiency of reconnection will decrease. This scaling has been confirmed through a number numerical simulations \citep{borovsky2007reconnection,malakit2010scaling}, laboratory experiments \citep{yoo2014laboratory},  as well as measurements from spacecraft flying through reconnecting plasmas in space \citep{borovsky2006effect,walsh2014simultaneous,walsh2014plasmaspheric,wang2015dependence,walsh2021role,zou2021geospace}. Relevant for the current study, this is also observed for global systems such as Earth's magnetosphere and its interaction with the solar wind. Global simulations have shown that with sufficient mass-loading, local conditions at the magnetopause dominate, and the integrated reconnection rate over the dayside magnetopause is reduced \citep{zhang2017transition}.

\section{Artificial mass-loading} \label{sec:implement}
Since magnetic reconnection is the predominant mechanism through which energy is deposited into Earth's magnetosphere, the intensity of a geomagnetic storm will be decreased if reconnection can be suppressed. Here, it is shown that through applying sufficient artificial mass-loading over a wide local time along the dayside magnetopause, the efficiency of reconnection can be reduced to the point where the intensity of a geomagnetic storm and space weather impacts are significantly reduced. The authors call this approach a StormWall.

\subsection{Implementation}
Implementation of this mass-loading could occur through a set of Earth-orbiting spacecraft, each holding a storage canister with mass-loading material. The material release would occur through a command sent from the ground when a user wished to reduce the impact of a solar wind structure such as an extreme coronal mass ejection. Once released from the storage canisters, the stored mass-loading material would rapidly photoionize to seed the magnetosphere with a plasma. Similar chemical releases have been conducted successfully by the space science community in the past using small canisters of barium and lithium to study plasma physics in Earth's magnetosphere \citep{krimigis1982active,hunton1998ionization}.

\begin{figure}[ht!]
\includegraphics[width=0.55\textwidth]{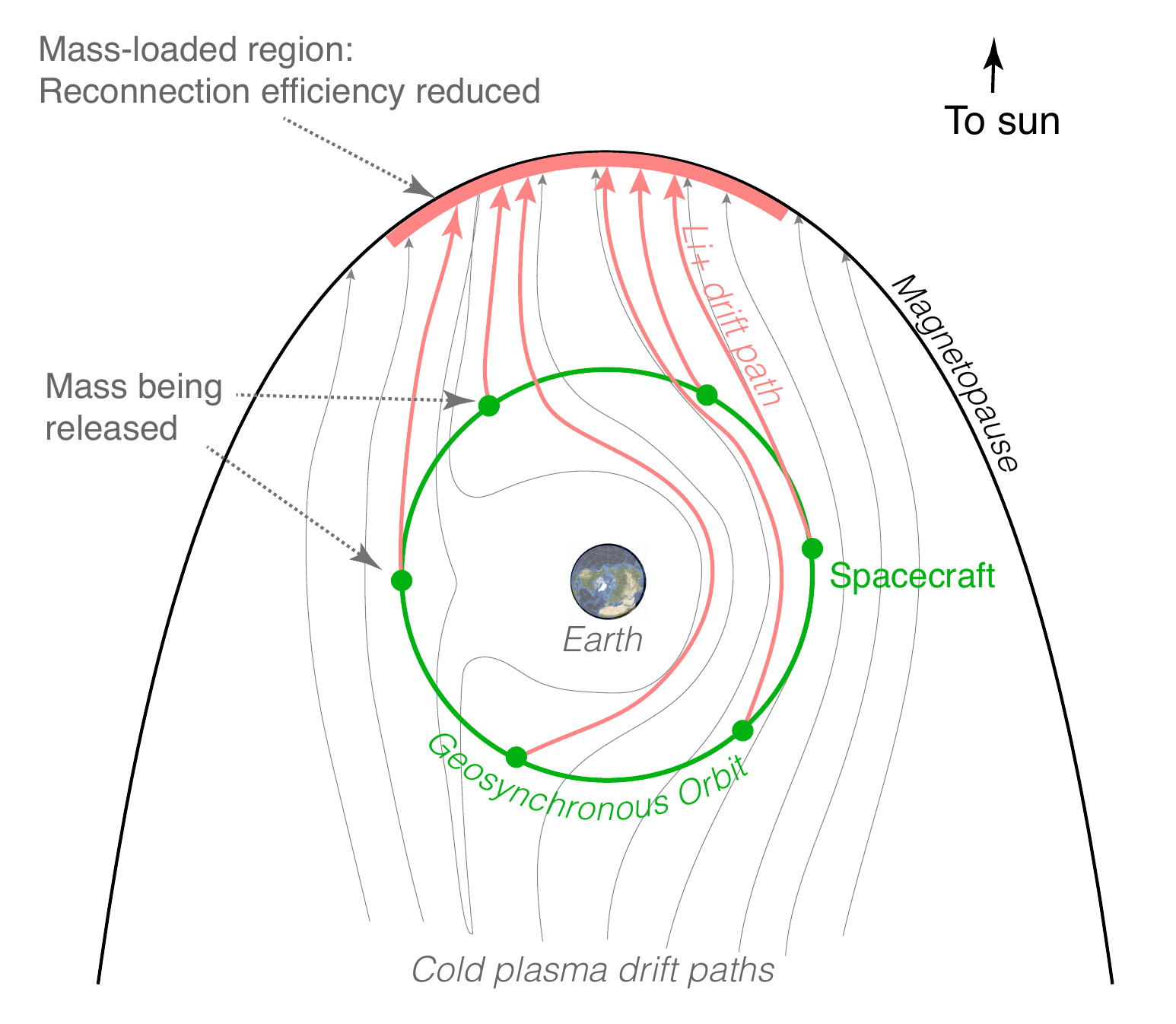}
\centering
\caption{Schematic diagram of mass-loading concept. Spacecraft in geosynchronous orbit release material that drifts to the dayside magnetopause. Geometry shows a slice in Earth's equatorialk plane. \label{fig:diagram}}
\end{figure}

The Earth-orbiting constellation could be flown in a circular geosynchronous orbit (\textit{R} = 6.6 $R_E$). Geosynchronous is a reasonable location since the spacecraft would be outside the Alfv\'en layer throughout the orbit during moderate to strong geomagnetic disturbances. Since the Alfv\'en layer marks the boundary between open and closed $E \times B$ drift paths, if the spacecraft releases a cool plasma, it will naturally drift to the dayside magnetopause to mass load the reconnecting boundary. Through releasing material at different local times along the geosynchronous orbit, the plasma will follow different drift paths and contact the dayside magnetopause at different local times. The drift time from geosynchronous to the magnetopause can range from 0.5 hrs to 2-3 hrs. Figure \ref{fig:diagram} presents a schematic diagram of this implementation. Six spacecraft are shown for example. The number could be greater or fewer depending on the implementation.

The protection provided by the mass-loading approach mimics an automobile airbag: installed once; ready to deploy at a moment’s notice; and requiring little maintenance. The constellation and mass-loading tanks could be launched in phases and stored in orbit for years. It would not be necessary to conduct rapid launches in real-time in response to an impending or occurring geomagnetic storm. Long-term maintenance of the constellation is low, requiring only minimal upkeep and new launches to only to replace spacecraft after mass release. When needed, the constellation would provide powerful, on-demand protection against extreme storms.

\subsection{Simulated Performance}
Numerical simulations were conducted to test the proposed model using the Space Weather Modeling Framework (SWMF) configured for geospace. SWMF Geospace couples the global Block Adaptive Tree Solar-wind Roe-type Upwind Scheme (BATS-R-US) magnetohydrodynamic (MHD) model \citep{powell1999solution,de2002adaptive, gombosi2021sustained} with a height-integrated ionosphere solver \citep{Ridley2004} and a kinetic ring current model \citep{DeZeeuw2004}. 
The configuration of the models follows the same as is used by NOAA's Space Weather Prediction Center (SWPC) for space weather monitoring and short-term forecasting and has a long history of reproducing key magnetosphere-ionosphere behavior \citep{Walsh2018a,Gordeev2015, Haiducek2017,Welling2019}.

To evaluate the feasibility of the physics in this technique, simulations were conducted of the major geomagnetic storm which occurred on 10-11 May 2024, or the ``Gannon" storm. This was selected as it is the largest storm in which upstream solar wind measurements are available to serve as inputs for the simulations. It represents a roughly 1 in 20 year storm. Two simulations were conducted, a reference run with no artificial mass-loading, and a second which included mass-loading. Solar wind measurements from NASA's WIND spacecraft were used as upstream inputs and ballistically propagated from the observation point to the model's upstream boundary.

\begin{figure}[ht!]
\includegraphics[width=0.9\textwidth]{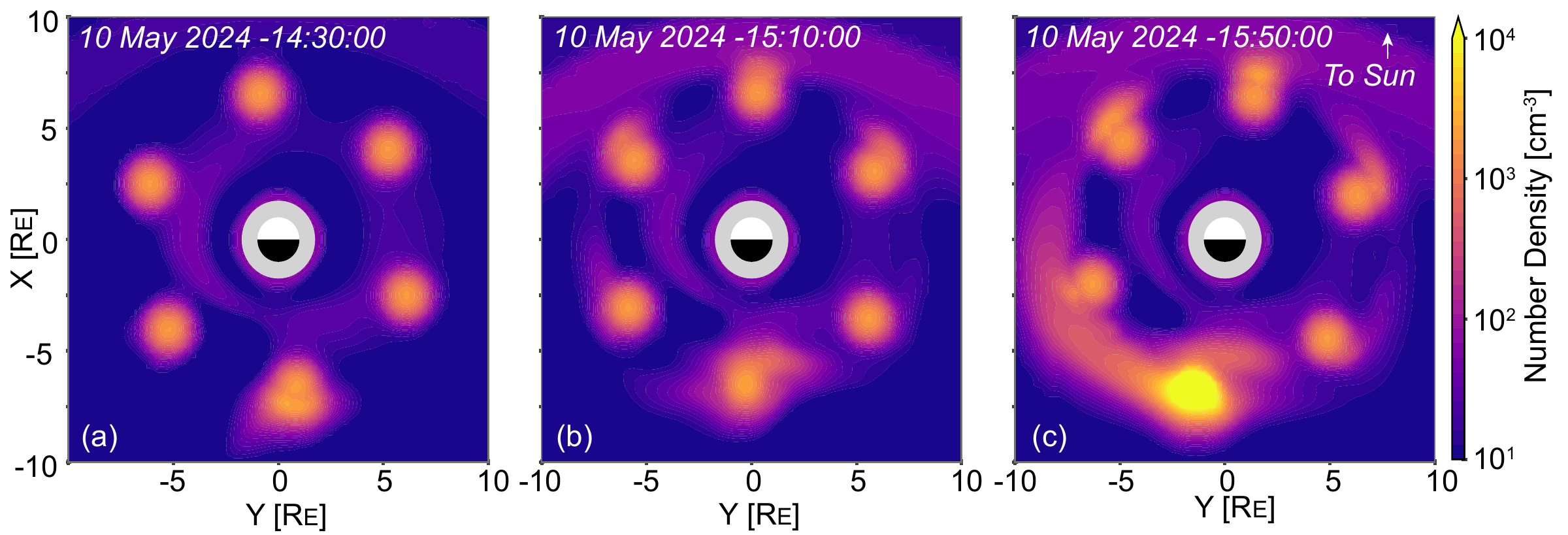}
\centering
\caption{Time-evolution of mass-loading process. Panels present simulated mass density in ecliptic plane shortly after mass-loading is initiated at the six spacecraft (panel (a)) and then at 40 minute intervals ((b) to (c)). X and Y axis are geocentric solar magnetospheric (GSM) coordinates in units of Earth radii.
\label{fig:xy_cuts}}
\end{figure}

Artificial mass loading was achieved by updating the BATS-R-US model such that it can include single-point mass sources that can represent a spacecraft releasing mass-loading material. Six moving geosynchronous spacecraft were included and turned ``on" for a total of 14 hours from 14:00 UT, 10 May 2024 to 4:00 UT, 11 May 2024. Each source was centered on the spacecraft with a central point that falls off as a function of radial distance using a Gaussian curve. This simulates the cloud nature that would form as the neutral material expands and is photoionized. During this mass-loading period, each spacecraft was releasing mass at a rate of 1.27 kg/s. The source terms have zero velocity and very low temperature (1 eV). Figure \ref{fig:xy_cuts} presents mass density in the simulation at three epochs: just after the sources are turned on and then at 40 minute intervals. Plasma drifts from the spacecraft towards the dayside magnetopause are visible in the density contours.

\begin{figure}[ht!]
\includegraphics[width=0.9\textwidth]{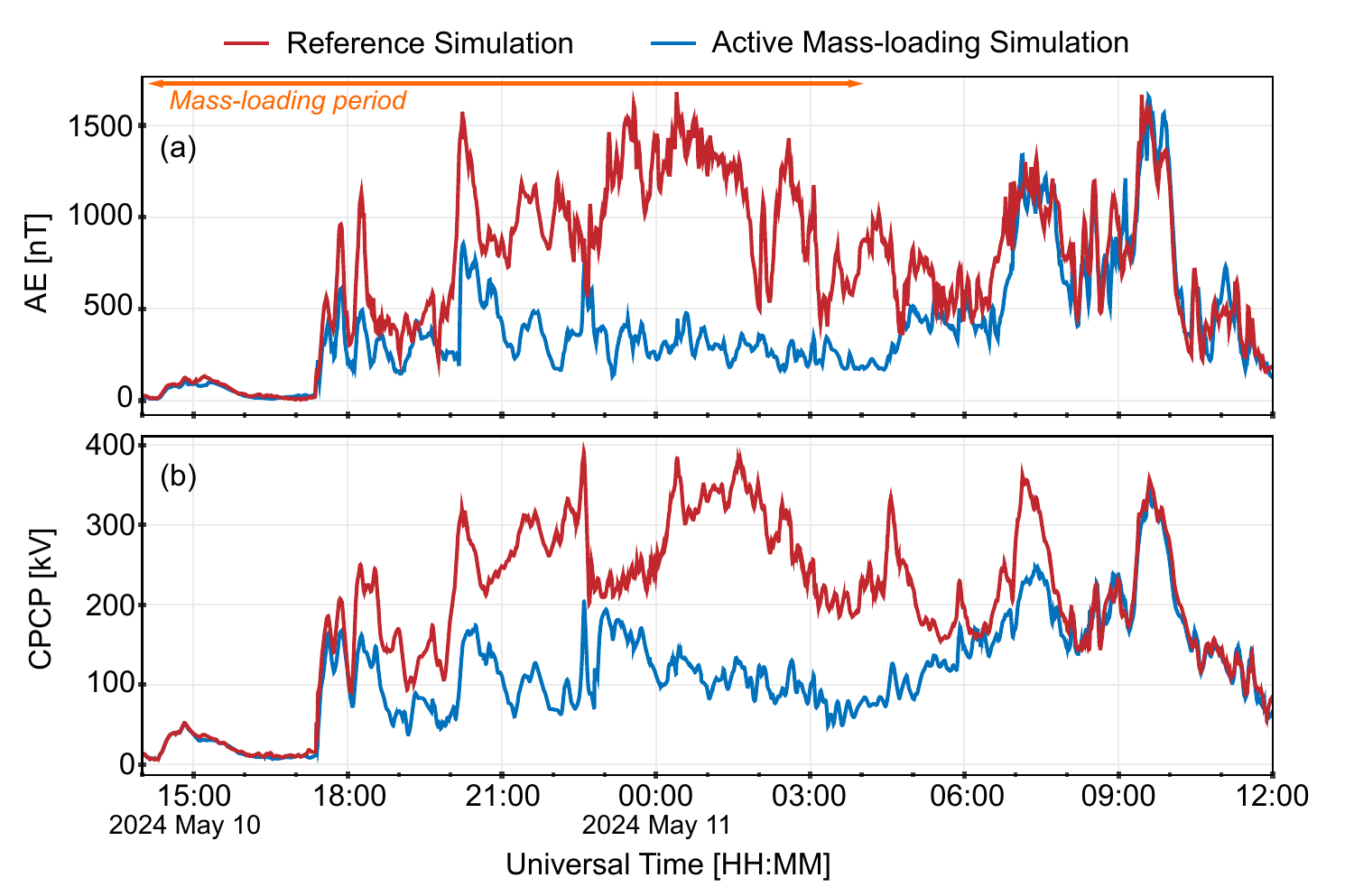}
\centering
\caption{Quantified impact of artificial mass-loading at the global level. Two measures of the intensity of geomagnetic storms are extracted from the model and presented, AE Index (a) and cross polar cap potential (b). The panels show the state of Earth's space-environment with and without mass-loading. The orange line shows the time window in which mass-loading is occurring (14:00 UT on 10 May 2024 to 4:00 UT on 11 May 2024). Mass-loading significantly reduces the impact of the CME on Earth's space environment through both measures of geomagnetic activity \label{fig:ae_cpcp}}
\end{figure}

The response of the model was evaluated to determine the impact of artificial mass-loading. Figure \ref{fig:ae_cpcp} presents two indices commonly used to determine space weather effects, the Auroral Electrojet (AE) index and the Cross Polar Cap Potential (CPCP). AE provides a measurement of the high latitude magnetic field perturbations on the surface of Earth which provides a measure of Geomagnetically Induced Current (GIC), while CPCP is a measure of convection in the ionosphere and solar wind magnetospheric coupling. For both parameters, a large and continuous reduction in intensity occurs. At the most intense period of the storm, near 0:30 UT on 11 May 2024, the reference simulation measures an AE index of 1600 nT, while the simulation with mass-loading experiences an AE of less than 250 nT, a reduction factor of more than a 84\%. At this same time, the reference simulation had a CPCP of 370 kV while the mass-loading simulation CPCP was 145 kV, a 61\% reduction. Mass-loading was turned off at 4:00 UT on 11 May 2024 in the simulation. Roughly two hours after mass-loading was turned off, the AE index and CPCP of the test and reference simulations were similar, indicating a time constant of the material in Earth's magnetospheric system.

\begin{figure}[ht!]
\includegraphics[width=0.9\textwidth]{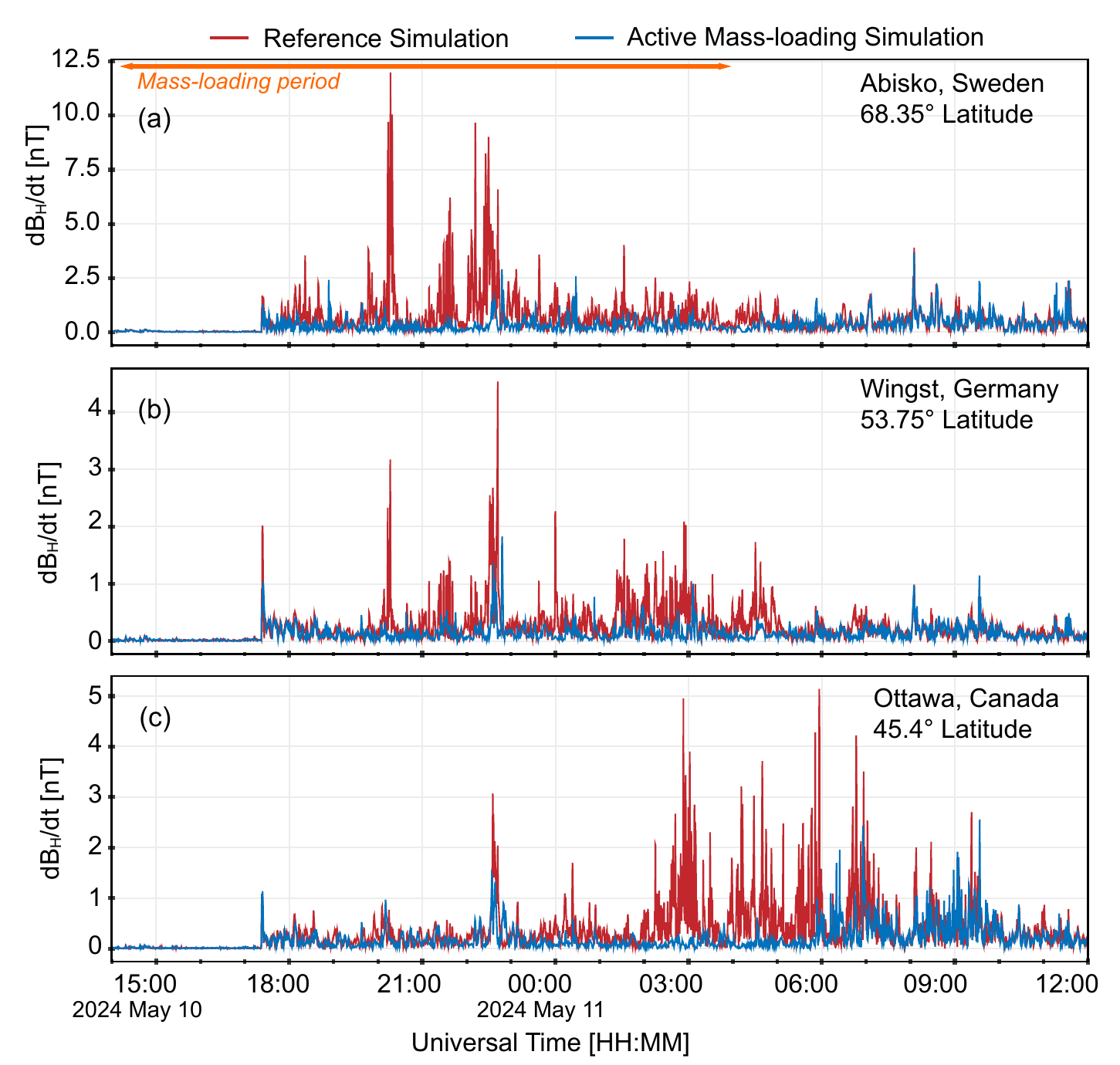}
\centering
\caption{Quantified impact of artificial mass-loading at the surface of Earth from the MHD simulation. $dB_{H}/dt$ is a measure of the change in the horizontal component in the magnetic field.}
\label{fig:dbdt}
\end{figure}

Figure \ref{fig:dbdt} further evaluates the impact of artificial mass-loading but quantifying the reduction in ground-level $dB/dt$. 
The rate of change of the ground magnetic field is directly related to Geomagnetically Induced Currents (GIC), which disrupt the high voltage power transmission network \citep{pulkkinen2017geomagnetically}.
The ground magnetic disturbance is calculated from the SWMF results using a series of Biot-Savart integrals covering the most important magnetospheric and auroral current systems \citep{Yu2010b}.
Here, the horizontal component is shown, following the practice in \cite{Pulkkinen2013}.
In each frame of Figure \ref{fig:dbdt}, the predicted $dB/dt$ is shown for the reference simulation and the simulation with artificial mass-loading (red and blue lines, respectively). Each frame presents a different location corresponding to a real-world magnetic observatory. Each station is responding to local auroral activity. When the mass loading is active, $dB/dt$ is drastically reduced, greatly lowering the space weather risk exposure to the power grid.

\section{Feasibility}
Active mass-loading during a realistic geomagnetic storm is feasible with modern technology. Here a sample implementation is presented to demonstrate feasibility.

\subsection{Launch capacity}
The mass-loading material would come from Earth and be transported to a storage location in geosynchronous orbit. The total mass required to orbit would be (1) the mass-loading material ($m_{ml}$), (2) a canister to hold the mass-loading material ($m_{cml}$), and (3) a spacecraft to control the system ($m_{sc}$). The sum of these elements is labeled the payload mass ($m_{pl} = m_{ml} + m_{cml} + m_{sc}$).

The first element is the mass-loading material. Over the course of the simulated geomagnetic storm, a total of 384,048 kg of mass-loading material was released from the spacecraft in the MHD simulation ($m_{ml}$). This value ($m_{ml}$ = 384,048 kg) will be used for the current calculation. Next, the mass of the canister is calculated. A typical structural mass-ratio or ``propellant mass fraction'' \textit{(mass of fuel)/(mass of fuel+structure)} for a rocket is 0.8 to 0.9. Here a similar ratio is assumes to calculate the mass of the canister used to hold the mass-loading material. Using 0.9, this results in $m_{cml}$ = 38,405 km. An important element of this activity is that the mass-loading payload does not need to be a single monolithic structure. It could be divided into smaller units or multiple launches if necessary. Lastly, the mass of the spacecraft is added. The simulation used 6 spacecraft. The spacecraft bus from the recent GOES-R mission in geosynchronous orbit is used to model the spacecraft mass (LM2100: 2,300 kg). With 6 spacecraft, the total mass for spacecraft buses is $m_{sc}$ = 13,800 kg. Bringing parts together, the total payload mass required in orbit for the complete protection system is then $m_{pl}$ = 436,253 kg.

The total required payload mass to orbit is within near-future technological capabilities of humans. The SpaceX Starship is anticipated to have a capacity to Mars of more than 100 metric tons or 100,000 kg when utilizing refueling in LEO \citep{starship20}. Although an anticipated mass capacity to geosynchronous is not provided by SpaceX, a simple conservative approximation can be made based on energy. Since a transfer from an initial 500 km circular Earth-orbit to Mars requires $\Delta V_{Mars}$ =5.59 km/s \citep{maiwald2024feasibility}, while a transfer from the same initial coplanar circular Earth-orbit to GEO requires less energy, $\Delta V_{GEO}$ = 3.82 km/s, it is assumed Starship's capacity to GEO is at least 100 metric tons or 100,000 kg. Assuming each spacecraft and mass-loading tank pair are carried by a single launch vehicle, this would result in 72,708 kg per vehicle and a total of 6 primary launches required. SpaceX leadership has recently stated the company expects to launch Starship every two weeks in the coming years \citep{Shotwell2024_StarshipCadence}, so the total time duration to conduct these launches could be less than two months. Although still in development, the Long March-9 vehicle is planned to have similar launch capacity.

An alternative to a geosynchronous orbit could be to fly the spacecraft in a geosynchronous transfer orbit. In this model, fewer launch resources would be required, and the spacecraft would still spend much of its time near apogee or geosynchronous orbit. Further, rather than 6, a larger number of smaller spacecraft could be deployed to utilize smaller launch vehicles. This approach would also apply well to international collaboration where different nations could provide different nodes in the global defense system.

\subsection{Mass-loading material}
A number of chemicals could be used to form the mass-loading material. The material is required to be stable during storage, have a low vaporization temperature, and to photoionize after it is released from the spacecraft, so it can turn into a plasma and follow plasma drifts to the dayside magnetopause. The low vaporization temperature is needed to allow the material to be stored as a solid and then be vaporized through standard chemical reactions before being deployed into space. A reasonable family of materials that could meet these requirements are alkaline Earth metals such as sodium, barium, calcium, or lithium or compounds including components of them.

This activity has been demonstrated through successful smaller-scale chemical releases using these materials in the past on a number of space-based platforms \citep{johnson1992combined,stenbaek1993observed,larsen2002winds}. One of the early orbital demonstrations flew on the Active Magnetospheric Particle Tracer Explorers (AMPTE)-Ion Release Module (IRM) spacecraft which launched in 1984 and carried 16 chemical release canisters, eight lithium and eight barium \citep{krimigis1982active,hausler1985ampte,haerendell}. In the IRM experiment, the barium and lithium material was stored as solid compounds and then heated and vaporized through a controlled thermite burn. The vaporized neutral material was ejected into space where it photoionized.

Although lithium is used as an example to show feasibility. Further work could be conducted to optimize chemical compounds for storage and release of massloading material. In previous applications, experimenters have prioritized materials that either would not typically occur naturally in Earth's magnetosphere or would emit light so that the resulting ions could be tracked.  These requirements could be relaxed for the current application, which may open more possibilities.

\section{Summary}\label{sec:conc}
Using physics-based numerical modeling, this work has shown that humans have the ability and technology to actively stop or reduce the intensity of a geomagnetic storm. Since magnetic reconnection at Earth's magnetopause is the primary process in which energy can be transferred, this study presents a method to reduce the global efficiency of reconnection, which in turn reduces the intensity of a geomagnetic storm. This is done through releasing material in space at geosynchronous orbit which then photoionizes and follows plasma drift paths to mass-load the dayside magnetopause. The supporting numerical simulations find significant reductions in a number of metrics used to quantify the amplitude of geomagnetic storms including CPCP, AE Index, and ground-based dB/dt measurements. This approach is analogous to humans building a stormwall to fortify a city against a flooding river. Similarly, this approach is labeled here as a StormWall. The total mass required is within the ability of current and near-future launch technologies and the process lends itself well to international collaboration. The threat of the space environment on human life and technology remains a major risk internationally. Response thus far has been primarily based on developing prediction systems. Here, rather than prediction, a method is provided for defense.

\backmatter





\bmhead{Competing interests}
The authors declare that they have no competing interests.

\bmhead{Acknowledgments}
Dr. Walsh acknowledges support from the NSF grant number 1845151.  Dr. Welling's effort was supported by the Solar Tsunamis Endeavor Programme, led by the University of Otago and funded by the New Zealand Ministry for Business, Innovation \& Employment.

\bibliography{sw}

\begin{thebibliography}{37}
\providecommand{\natexlab}[1]{#1}
\providecommand{\url}[1]{{#1}}
\providecommand{\urlprefix}{URL }
\providecommand{\doi}[1]{\url{https://doi.org/#1}}
\providecommand{\eprint}[2][]{\url{#2}}
 \bibcommenthead

\bibitem[{Schulte in~den B{\"a}umen et~al(2014)Schulte in~den B{\"a}umen, Moran, Lenzen, Cairns, and Steenge}]{schulte2014severe}
Schulte in~den B{\"a}umen H, Moran D, Lenzen M, et~al (2014) How severe space weather can disrupt global supply chains. Natural Hazards and Earth System Sciences 14(10):2749--2759

\bibitem[{Berger(2024)}]{Shotwell2024_StarshipCadence}
Berger E (2024) Spacex president predicts rapid increase in starship launch rate. \urlprefix\url{https://arstechnica.com/space/2024/11/spacex-president-predicts-rapid-increase-in-starship-launch-rate/}

\bibitem[{Borovsky and Denton(2006)}]{borovsky2006effect}
Borovsky JE, Denton MH (2006) Effect of plasmaspheric drainage plumes on solar-wind/magnetosphere coupling. Geophysical Research Letters 33(20)

\bibitem[{Borovsky and Hesse(2007)}]{borovsky2007reconnection}
Borovsky JE, Hesse M (2007) The reconnection of magnetic fields between plasmas with different densities: Scaling relations. Physics of Plasmas 14(10):102309

\bibitem[{Cassak and Shay(2007)}]{cassak2007scaling}
Cassak P, Shay M (2007) Scaling of asymmetric magnetic reconnection: General theory and collisional simulations. Physics of Plasmas 14(10):102114

\bibitem[{De~Zeeuw et~al(2002)De~Zeeuw, Gombosi, Groth, Powell, and Stout}]{de2002adaptive}
De~Zeeuw DL, Gombosi TI, Groth CP, et~al (2002) An adaptive mhd method for global space weather simulations. IEEE Transactions on Plasma Science 28(6):1956--1965

\bibitem[{De~Zeeuw et~al(2004)De~Zeeuw, Sazykin, Wolf, Gombosi, Ridley, and Tóth}]{DeZeeuw2004}
De~Zeeuw DL, Sazykin S, Wolf RA, et~al (2004) Coupling of a global {{MHD}} code and an inner magnetospheric model: {{Initial}} results 109(A12):A12219. \doi{10.1029/2003JA010366}, \urlprefix\url{http://doi.wiley.com/10.1029/2003JA010366}

\bibitem[{Gombosi et~al(2021)Gombosi, Chen, Glocer, Huang, Jia, Liemohn, Manchester, Pulkkinen, Sachdeva, Al~Shidi et~al}]{gombosi2021sustained}
Gombosi TI, Chen Y, Glocer A, et~al (2021) What sustained multi-disciplinary research can achieve: The space weather modeling framework. Journal of Space Weather and Space Climate 11:42

\bibitem[{Gordeev et~al(2015)Gordeev, Sergeev, Honkonen, Kuznetsova, Rastätter, Palmroth, Janhunen, Tóth, Lyon, and Wiltberger}]{Gordeev2015}
Gordeev E, Sergeev V, Honkonen I, et~al (2015) Assessing the performance of community-available global {{MHD}} models using key system parameters and empirical relationships 13(12):868--884. \doi{10.1002/2015SW001307}, \urlprefix\url{http://doi.wiley.com/10.1002/2015SW001307}

\bibitem[{Haerendel et~al(1985)Haerendel, Valenzuela, F{\"o}ppl, Rieger, and Bauer}]{haerendell}
Haerendel G, Valenzuela A, F{\"o}ppl H, et~al (1985) The li/ba release experiments on board the ampte irm satellite. IEEE Trans Geosci Remote Sensing pp 253--258

\bibitem[{Haiducek et~al(2017)Haiducek, Welling, Ganushkina, Morley, and Ozturk}]{Haiducek2017}
Haiducek JD, Welling DT, Ganushkina NY, et~al (2017) {{SWMF Global Magnetosphere Simulations}} of {{January}} 2005: {{Geomagnetic Indices}} and {{Cross-Polar Cap Potential}} 15(12):1567--1587. \doi{10.1002/2017SW001695}, \urlprefix\url{http://doi.wiley.com/10.1002/2017SW001695}

\bibitem[{Hausler et~al(1985)Hausler, Melzner, Stocker, Valenzuela, Bauer, Parigger, Sigritz, Schoning, Seidenschwang, Eberl et~al}]{hausler1985ampte}
Hausler B, Melzner F, Stocker J, et~al (1985) The ampte irm spacecraft. IEEE transactions on geoscience and remote sensing (3):192--201

\bibitem[{Hunton et~al(1998)Hunton, Wolf, and Shadid}]{hunton1998ionization}
Hunton D, Wolf P, Shadid T (1998) Ionization mechanisms in crres chemical releases: 1. in situ measurements and model results. Journal of Geophysical Research: Space Physics 103(A1):457--470

\bibitem[{Johnson and Kierein(1992)}]{johnson1992combined}
Johnson M, Kierein J (1992) Combined release and radiation effects satellite (crres): Spacecraft and mission. Journal of Spacecraft and Rockets 29(4):556--563

\bibitem[{Krimigis et~al(1982)Krimigis, Haerendel, McEntire, Paschmann, and Bryant}]{krimigis1982active}
Krimigis S, Haerendel G, McEntire R, et~al (1982) The active magnetospheric particle tracer explorers (ampte) program. Eos, Transactions American Geophysical Union 63(45):843--850

\bibitem[{Larsen(2002)}]{larsen2002winds}
Larsen MF (2002) Winds and shears in the mesosphere and lower thermosphere: Results from four decades of chemical release wind measurements. Journal of Geophysical Research: Space Physics 107(A8):SIA--28

\bibitem[{Lingam and Loeb(2017)}]{lingam2017risks}
Lingam M, Loeb A (2017) Risks for life on habitable planets from superflares of their host stars. The Astrophysical Journal 848(1):41

\bibitem[{Loper(2019)}]{loper2019carrington}
Loper RD (2019) Carrington-class events as a great filter for electronic civilizations in the drake equation. Publications of the Astronomical Society of the Pacific 131(998):044202

\bibitem[{Maiwald et~al(2024)Maiwald, Bauerfeind, F{\"a}lker, Westphal, and Bach}]{maiwald2024feasibility}
Maiwald V, Bauerfeind M, F{\"a}lker S, et~al (2024) About feasibility of spacex's human exploration mars mission scenario with starship. Scientific Reports 14(1):11804

\bibitem[{Malakit et~al(2010)Malakit, Shay, Cassak, and Bard}]{malakit2010scaling}
Malakit K, Shay M, Cassak P, et~al (2010) Scaling of asymmetric magnetic reconnection: Kinetic particle-in-cell simulations. Journal of Geophysical Research: Space Physics 115(A10)

\bibitem[{Powell et~al(1999)Powell, Roe, Linde, Gombosi, and De~Zeeuw}]{powell1999solution}
Powell KG, Roe PL, Linde TJ, et~al (1999) A solution-adaptive upwind scheme for ideal magnetohydrodynamics. Journal of Computational Physics 154(2):284--309

\bibitem[{Pulkkinen et~al(2013)Pulkkinen, Rastätter, Kuznetsova, Singer, Balch, Weimer, Toth, Ridley, Gombosi, Wiltberger, Raeder, and Weigel}]{Pulkkinen2013}
Pulkkinen A, Rastätter L, Kuznetsova M, et~al (2013) Community-wide validation of geospace model ground magnetic field perturbation predictions to support model transition to operations. Space Weather 11(6):369--385. \doi{10.1002/swe.20056}, \urlprefix\url{http://doi.wiley.com/10.1002/swe.20056}

\bibitem[{Pulkkinen et~al(2017)Pulkkinen, Bernabeu, Thomson, Viljanen, Pirjola, Boteler, Eichner, Cilliers, Welling, Savani et~al}]{pulkkinen2017geomagnetically}
Pulkkinen A, Bernabeu E, Thomson A, et~al (2017) Geomagnetically induced currents: Science, engineering, and applications readiness. Space weather 15(7):828--856

\bibitem[{Ridley et~al(2004)Ridley, Gombosi, and DeZeeuw}]{Ridley2004}
Ridley AJ, Gombosi TI, DeZeeuw DL (2004) Ionospheric control of the magnetosphere: Conductance 22(2):567--584. \doi{10.5194/angeo-22-567-2004}, \urlprefix\url{http://www.ann-geophys.net/22/567/2004/}

\bibitem[{SpaceX(2020)}]{starship20}
SpaceX (2020) Starship users guide. \urlprefix\url{https://spacex.com.pl/files/2020-04/starship-users-guide-v1.pdf}

\bibitem[{Stenbaek-Nielsen et~al(1993)Stenbaek-Nielsen, Wescott, and Hallinan}]{stenbaek1993observed}
Stenbaek-Nielsen H, Wescott E, Hallinan T (1993) Observed barium emission rates. Journal of Geophysical Research: Space Physics 98(A10):17491--17500

\bibitem[{Walsh et~al(2014{\natexlab{a}})Walsh, Foster, Erickson, and Sibeck}]{walsh2014simultaneous}
Walsh B, Foster J, Erickson P, et~al (2014{\natexlab{a}}) Simultaneous ground-and space-based observations of the plasmaspheric plume and reconnection. Science 343(6175):1122--1125

\bibitem[{Walsh et~al(2014{\natexlab{b}})Walsh, Phan, Sibeck, and Souza}]{walsh2014plasmaspheric}
Walsh B, Phan T, Sibeck D, et~al (2014{\natexlab{b}}) The plasmaspheric plume and magnetopause reconnection. Geophysical Research Letters 41(2):223--228

\bibitem[{Walsh and Zou(2021)}]{walsh2021role}
Walsh BM, Zou Y (2021) The role of magnetospheric plasma in solar wind-magnetosphere coupling: A review. Journal of Atmospheric and Solar-Terrestrial Physics 219:105644

\bibitem[{Walsh et~al(2018)Walsh, Welling, Zou, and Nishimura}]{Walsh2018a}
Walsh BM, Welling DT, Zou Y, et~al (2018) A {{Maximum Spreading Speed}} for {{Magnetopause Reconnection}} 45(11):5268--5273. \doi{10.1029/2018GL078230}, \urlprefix\url{http://doi.wiley.com/10.1029/2018GL078230}

\bibitem[{Wang et~al(2015)Wang, Kistler, Mouikis, and Petrinec}]{wang2015dependence}
Wang S, Kistler LM, Mouikis CG, et~al (2015) Dependence of the dayside magnetopause reconnection rate on local conditions. Journal of Geophysical Research: Space Physics 120(8):6386--6408

\bibitem[{Weaver et~al(2004)Weaver, Murtagh, Balch, Biesecker, Combs, Crown, Doggett, Kunches, Singer, and Zezula}]{weaver2004halloween}
Weaver M, Murtagh W, Balch C, et~al (2004) Halloween space weather storms of 2003 noaa technical memorandum oar sec-88. NOAA and Space Environment Center, Colorado, USA

\bibitem[{Welling(2019)}]{Welling2019}
Welling D (2019) Magnetohydrodynamic {{Models}} of {{B}} and {{Their Use}} in {{GIC Estimates}}. In: Gannon JL, Swidinsky A, Xu. Z (eds) Geomagnetically {{Induced Currents}} from the {{Sun}} to the {{Power Grid}}. American Geophysical Union (AGU), p 43--65, \doi{10.1002/9781119434412.ch3}, \urlprefix\url{https://onlinelibrary.wiley.com/doi/abs/10.1002/9781119434412.ch3}

\bibitem[{Yoo et~al(2014)Yoo, Yamada, Ji, Jara-Almonte, Myers, and Chen}]{yoo2014laboratory}
Yoo J, Yamada M, Ji H, et~al (2014) Laboratory study of magnetic reconnection with a density asymmetry across the current sheet. Physical review letters 113(9):095002

\bibitem[{Yu et~al(2010)Yu, Ridley, Welling, and Tóth}]{Yu2010b}
Yu Y, Ridley AJ, Welling DT, et~al (2010) Including gap region field-aligned currents and magnetospheric currents in the {MHD} calculation of ground-based magnetic field perturbations. Journal of Geophysical Research: Space Physics 115(8). \doi{10.1029/2009JA014869}

\bibitem[{Zhang et~al(2017)Zhang, Brambles, Cassak, Ouellette, Wiltberger, Lotko, and Lyon}]{zhang2017transition}
Zhang B, Brambles O, Cassak P, et~al (2017) Transition from global to local control of dayside reconnection from ionospheric-sourced mass loading. Journal of Geophysical Research: Space Physics 122(9):9474--9488

\bibitem[{Zou et~al(2021)Zou, Walsh, Shi, Lyons, Liu, Angelopoulos, Ruohoniemi, Coster, and Henderson}]{zou2021geospace}
Zou Y, Walsh BM, Shi X, et~al (2021) Geospace plume and its impact on dayside magnetopause reconnection rate. Journal of Geophysical Research: Space Physics 126(6):e2021JA029117

\end{thebibliography}

\end{document}